\documentclass[namedreferences]{kluwer}

\usepackage{graphicx}
\def\pmb#1{\setbox0=\hbox{$#1$}%
    \kern-.025em\copy0\kern-\wd0\kern.05em\copy0\kern-\wd0\kern-.025em
    \raise.0433em\box0}

\begin{document}
\begin{article}
\begin{opening}
\title{Some aspects of Relativistic Astrometry from within the Solar
    System}
\author{M.~T.~\surname{Crosta} \email{crosta@to.astro.it}}
\author{A.~\surname{Vecchiato} \email{vecchiato@to.astro.it}}
\institute{University of Padua and INAF - Observatory of Turin, Italy}
\author{F.~\surname{de Felice} \email{defelice@pd.infn.it}}
\institute{University of Padua, Italy}
\author{M.~G.~\surname{Lattanzi} \email{lattanzi@to.astro.it}}
\author{B.~\surname{Bucciarelli} \email{bucciarelli@to.astro.it}}
\institute{INAF - Observatory of Turin, Italy}

\runningtitle{Some aspects of Relativistic Astrometry from within the
  Solar System}
\runningauthor{Crosta~M.~T. et al.}
\begin{ao}\\
Maria Teresa Crosta \\
INAF - Osservatorio Astronomico di Torino \\ 
strada Osservatorio 20 \\ 
10025 Pino Torinese (TO) \\
Italy 
\end{ao}

\begin{abstract}
  In this article we outline the structure of a general relativistic
  astrometric model which has been developed to deduce the position
  and proper motion of stars from 1-microarcsecond optical
  observations made by an astrometric satellite orbiting around the
  Sun.  The basic assumption of our model is that the Solar System is
  the only source of gravity, hence we show how we modeled the
  satellite observations in a many-body perturbative approach limiting
  ourselves to the order of accuracy of $(v/c)^2$. The microarcsecond
  observing scenario outlined is that for the GAIA astrometric
  mission.
\end{abstract}


\abbreviations{\abbrev{GR} {General Theory of Relativity};
               \abbrev{CM} {Center of Mass}}



\end{opening}

\section{Introduction}
Aim of fundamental astrometry is to measure stellar positions,
pa\-rallaxes and proper motion directly from optical observations.
Understanding how electromagnetic waves propagate in a time dependent
gravitational field is essential to answer basic questions in
astrophysics. In fact, most of the physical information about the
surrounding universe (see \opencite{1999PhRvD..59h4023K},
\opencite{2001ApJ...556L...1K} and references therein) is carried by
light signals. Nowadays the scientific objectives of astrometry and
astrophysics can be combined thanks to the precision achievable by the
new generation of space astrometry missions like GAIA of the European
Space Agency (ESA). GAIA will be launched not later then 2012 (with a
possible window in 2010) and is expected to make astrometric
measurements with a precision of a few $\mu$arcsec
\cite{2001AAp...369..339P}. At this level of accuracy one should
consider contributions to the light deflection not only by the mass of
the Sun and of the other planets, but also by their gravitational
quadrupole moments as well as their translational and rotational
motion (table~\ref{tab:releff}). This implies that a sui\-table
reduction of the GAIA observational data requires the development of a
model of the celestial sphere more sophisticated than the one used for
Hipparcos, the first global astrometry satellite launched by ESA in
1989.  All mathematical and physical assumptions which characterize
our model, as we shall specify in sections~\ref{sec:model}
and~\ref{sec:test}, are tailored to the GAIA concept.

\begin{table}
  \begin{tabular}{cr@{}lr@{}lr@{}lc} \hline
     & \multicolumn{2}{c}{$\delta\chi_\textrm{pN}$} &
       \multicolumn{2}{c}{$\delta\chi_{J_{2}}$} &
       \multicolumn{2}{c}{$\delta\chi_{L}$} &
       $\chi_\textrm{max}$ \\ \hline
     Sun      & $1.\!\!''75$ &          & $\sim 1$ & ~$\mu$as & 0.7 & ~$\mu$as & $(180^\circ)$ \\
     Mercury  &         $83$ & ~$\mu$as &   --  & &   --  & & $(7')$ \\
     Venus    &        $493$ &          &   --  & &   --  & & $(4.0^\circ )$ \\
     Earth    &        $574$ &          & $0.6$ & &   --  & & $(101^\circ)$ \\
     Moon     &         $26$ &          &   --  & &   --  & & $(2.3^\circ)$ \\
     Mars     &        $116$ &          & $0.2$ & &   --  & & $(17')$ \\ 
     Jupiter  &      $16290$ &          & $240$ & & $0.2$ & & $(87^\circ/3')$ \\
     Saturn   &       $5772$ &          &  $94$ & &   --  & & $(16^\circ/51'')$ \\ 
     Uranus   &       $2030$ &          &   $7$ & &   --  & & $(67'/4'')$ \\
     Neptune  &       $2487$ &          &   $8$ & &   --  & & $(50'/3'')$ \\ 
     Pluto    &          $7$ &          &   --  & &   --  & & $(0''\!\!.3)$ \\ 
     Io       &         $30$ &          &   --  & &   --  & & $(18'')$ \\
     Europe   &         $19$ &          &   --  & &   --  & & $(10'')$ \\ 
     Ganymede &         $34$ &          &   --  & &   --  & & $(30'')$ \\ 
     Callisto &         $27$ &          &   --  & &   --  & & $(22'')$ \\ \hline
  \end{tabular}
\caption{Detectable first order post-Newtonian relativistic effects in 
Solar System at 1~$\mu$arcsec level. Here $\delta\chi_{\mathrm{pN}}$ 
is the contribution due to the spherically symmetric component of the
gravitational field of the corresponding body, $J_2$ is that of the 
quadrupolar moment, and $L$ that of the rotational motion. The values 
whitin the parenthesis represents the maximum angular distance between 
the perturbing body and photon at which the effect still attains 
1~$\mu$arcsec; where two values are reported, they refer to the pN and 
the $J_2$ cases respectively. The pN values have been calculated 
considering an observer at L2 in the Sun-Earth system, and using the 
data reported in \protect\inlinecite{1999ssd.book.....M} for the Solar 
System bodies.}
\label{tab:releff}
\end{table}

GAIA's main scientific objective is to shed definitive light on the
origin, structure and kinematics of the Galaxy utilizing a complete
census of the entire sky down to the visual magnitude $V=20$, for a
total of about one billion stars. This survey is achieved by
exploiting the same scanning strategy which was used with the
successful astrometric mission Hipparcos.  In this way GAIA will
measure the effects of many thousands of extra solar planets, and
determine their orbit; thousands of brown dwarf and white dwarfs will
also be identified.  Many thousands of new minor bodies, inner
Trojans, and even new Trans-Neptune objects, including Plutinos, may
be discovered.  The main physical properties of asteroids ($10^5-10^6$
new objects) will be investigated including masses, densities, sizes,
shapes and taxonomic classes. In particular, asteroid mass is
determined by measuring the tiny gravitational perturbations
experienced in case of mutual close approaches
\cite{2002gesp.conf.....B}.

Moreover GAIA's $\mu$arcsecond global astrometry allows one to test
General Relativity (GR).  Realistic end-to-end simulations in fact
\cite{2003AAp...399..337V} show that GAIA could measure the PPN
parameter $\gamma$ to $\sim 10^{-7}$ (1$\sigma$) after 5 years of
continuous observations. The para\-meter $\gamma$ measures the
\textit{excess} of curvature produced by mass-energy as compared to GR
where its value is 1. As well known (\opencite{1993PhRvL..70.2217D};
\citeauthor{2002PhRvL..89h1601D}
\citeyear{2002PhRvL..89h1601D,2002PhRvD..66d6007D}) a deviation from
$\gamma$'s GR value would deeply affect our understanding of
fundamental physics.

In what follows, greek indices run from 0 to 3 and latin indices run
from 1 to 3.

\section{Modeling the observations of GAIA}
\label{sec:model}
The astrometric problem consists in the determination, from a
prescribed set of observables, of the astrometric parameters of a star
namely its coordinates, parallax and proper motion. While in classical
astrometry these quantities are well defined, in GR they must be
interpreted consistently with the relativistic framework of the model.

The first step of the modeling is to identify the gravitational
sources and then fix the background geometry. Next we choose a
suitable reference system to label space-time points and describe the
light trajectory, the stellar motion and the motion of the observer.
Evidently the goal of our model is to write a formula which relates
the observables to the astrometric parameters. To this purpose we had
to take into account the way how the satellite GAIA will operate.

Our first assumption is that the Solar System is the only source of
gravity; in doing so we ignore, e.g., microlensing effects\footnote{As
  far as the mission GAIA is concerned we expect $\sim 25000$
  potential microlensing events along the Galactic disk during the
  five year \cite{2002MNRAS.331..649B}.}  which are perhaps the most
important perturbations suffered by a light ray in its way to us since
they generate systematic errors in the data reduction.  However, their
expected number is very small as compared to the number of stars
surveyed by GAIA ($\ge 10^9$) hence we feel justified neglecting them
at least at this stage of modeling.

The second assumption is to consider the Solar System as a source of a
weak gravitational field.  This allows us to adopt a quasi-Minkowskian
metric
\begin{equation}
g_{\alpha \beta}= \eta_{\alpha\beta}+h_{\alpha\beta}+O(h^2)
\label{eq:metsum}
\end{equation}
where the $h_{\alpha\beta}$ are perturbation coefficients such that
$|h_{\alpha\beta}|\ll 1$; their spatial variations are at most of the
order of $|h_{\alpha\beta}|$ while their time variations are at most
of the order of $(v/c)|h_{\alpha\beta}|$ where $c$ is the velocity of
light in vacuum and $v$ is the typical velocity within the perturbing
system.  Notice that $v$ represents the velocity needed to be
gravitationally bounded in the Solar System
\cite{1987thyg.book.s6.9D}, and its typical value for the energy
balance (virial theorem) is $\sim 10~\textrm{km}\,\textrm{s}^{-1}$.
Under these conditions we shall take $h_{\alpha\beta}=\sum_a
h^{(a)}_{\alpha \beta}$ where the sum is extended to the bodies of the
Solar System. The order of magnitude of each term $h^{(a)}_{\alpha
  \beta}$ can be expressed in term of powers of $(v/c)$. Since the $h$
terms are at least of order $(v/c)^2$, the level of accuracy which is
expected to be reached within the model is fixed by the order of
$(v/c)$ to which one likes to extend the calculation.  Having these
considerations in mind, we shall adopt the IAU recommended metric form
for the barycentric reference frame inside the Solar System
\cite{2000IAU-res....B1.3}. In this case the lowest orders of
magnitude of the metric coefficients are $h_{00}=O((v/c)^2)$,
$h_{ij}=O((v/c)^2)$ and $h_{0i}=O((v/c)^3)$. In the following we
describe the construction of a model accurate to $(v/c)^2$ where in
addition we approximate the Solar System to a static, non-rotating and
non-expanding gravitating system and therefore we do not need to
consider the spatial location of the individual gravitational sources
at the corresponding retarded time.  This assumption implies that in
reconstructing the light trajectory of each photon emitted by a
distant star, we keep the bodies of the Solar System fixed at the
position they have with respect to its center of mass (CM) at the time
of observation, say. Evidently each subsequent light ray will be
reconstructed updating the positions of the bodies of the Solar System
according to their actual motion. This does not mean that the bodies
are at rest in the model, but that in our $(v/c)^2$ approximation
their position can be considered fixed in a single integration. This
model will serve as a touchstone for comparison once we extend our
analysis to the order of $(v/c)^3$.

\begin{figure}[htp]
\centerline{\includegraphics[width=8.5cm]{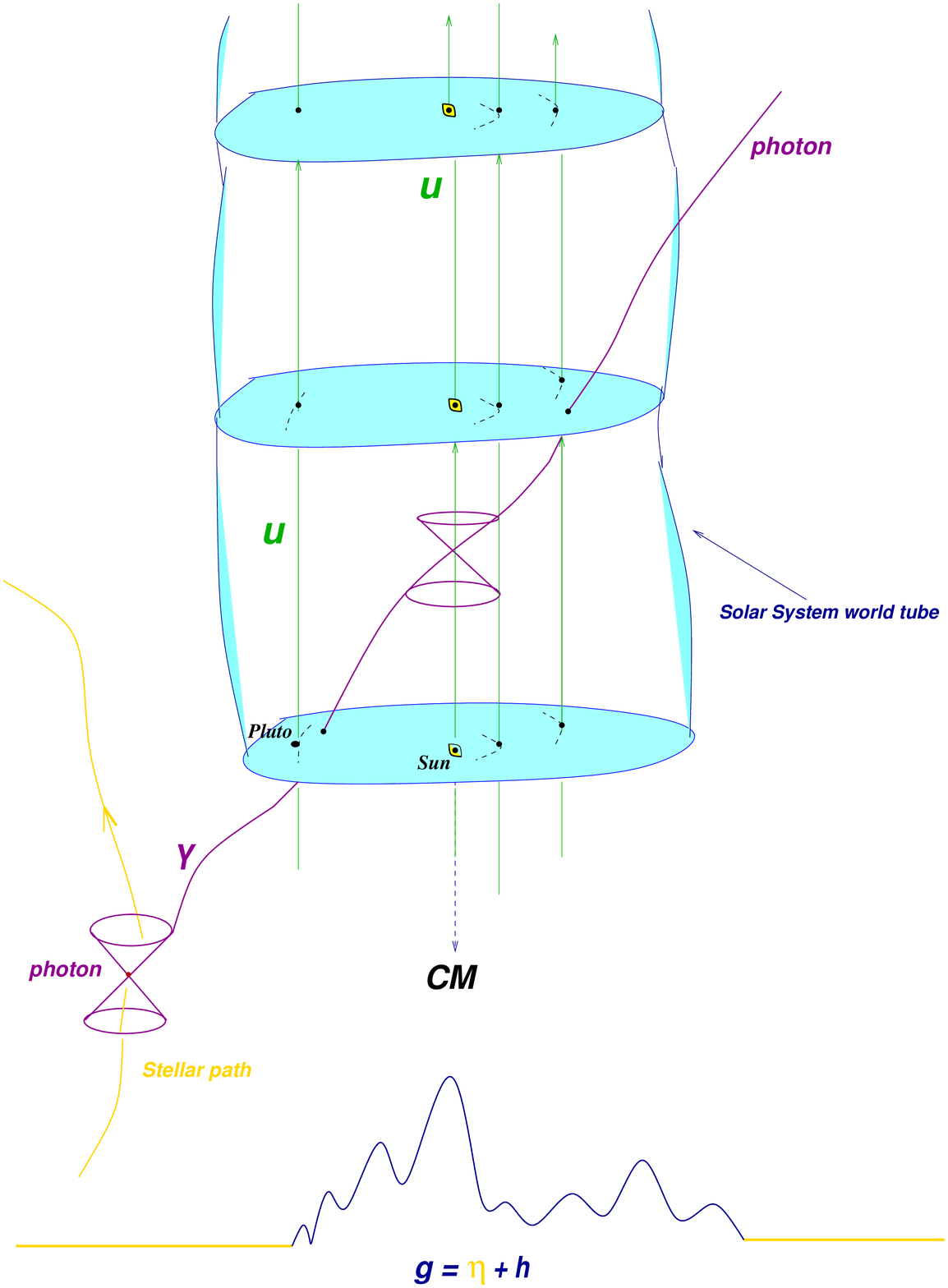}}
\caption{The astronomical set-up. A photon crosses the Solar System which
  is the only source of gravity. The vector field $\pmb{u}$ describes
  physical observers who Lie-transports the spatial coordinates
  (including those of the planets at the $(v/c)^2$ order).}
\label{fig:sswt}
\end{figure}

Let $(\tau, x,y,z)$ be a coordinate system with respect to which the
spacetime metric takes the form (\ref{eq:metsum}). Up to the order
$(v/c)^2$ the space-time certainly admits a vorticity free congruence
of curves which allows for a spacetime foliation into three
dimensional hypersurfaces with equation $\tau=$constant; we shall term
them $S(\tau)$ and identify the normals with a unitary vector field
$u^\alpha=dx^\alpha/d\sigma$ with $x^\alpha\equiv{\tau, x,y,z}$ as
$\alpha=0,1,2,3$. In this case the spatial coordinates can be fixed
within each slice up to spatial transformations only. We can then give
the vector field $u^\alpha$ the form
$u^\alpha=(-g_{00})^{-1/2}\delta^\alpha_\tau$ which makes the vector
unitary. The vector field so defined describes a physical observer
with proper time $\sigma$ who Lie-transports the spatial coordinates.
We also require that the CM of the Solar System
(figure~\ref{fig:sswt}) and the planets belong to this congruence. At
this point of the modeling we have characterized the metric, the
coordinate system and a physical observer $\pmb{u}$ who is at rest
with respect to the CM of the Solar System and therefore termed
baricentric; we can now trace the light path.

\begin{figure}
\centerline{\includegraphics[width=10cm]{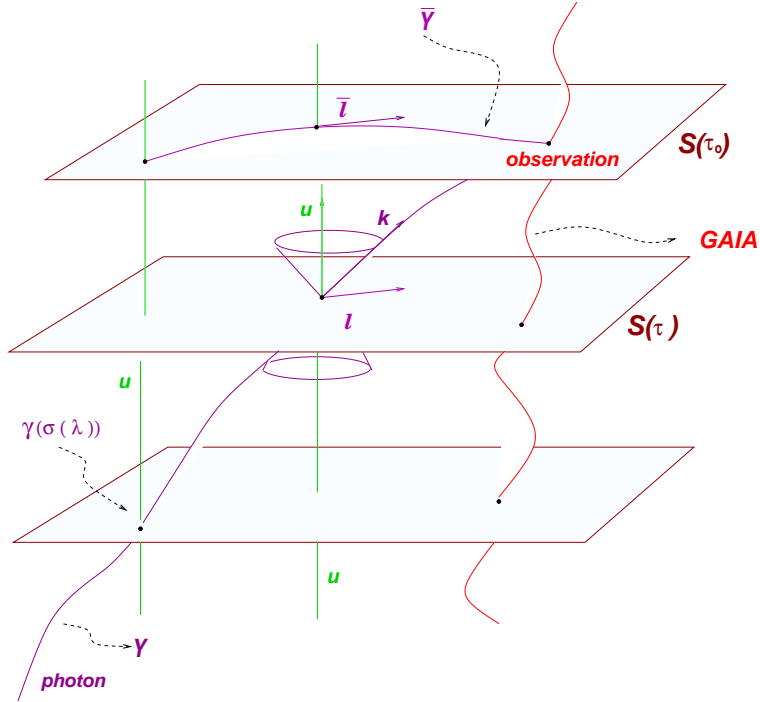}}
\caption{Mapping the null ray into the rest space of the barycentric
  observer in the case of a model accurate at the $(v/c)^2$ order.}
\label{fig:map}
\end{figure}

Let $\mit\Upsilon$ be a null geodesic with tangent vector field
$k^\alpha\equiv d\xi^\alpha/d\lambda$ (here $\lambda$ is a real
parameter on $\mit\Upsilon$). A photon traveling from a distant star
to the astrometric satellite within the Solar System would see the
spacetime as a time development of the spatial hypersurfaces
$S(\tau)$.  The light trajectory will end at the satellite at a
coordinate time $\tau_0$ and at a point with spatial coordinates
$x^i_o$.  Because the spacetime is foliated into slices $S(\tau)$ each
representing a surface of simultaneity for the baricentric observer
$\pmb{u}$ namely her/his rest space, each point
$\mit\Upsilon(\lambda)$ on the null geodesic can be mapped into the
point on the slice $S(\tau_0)$ where the unique normal to $S(\tau)$ at
$\mit\Upsilon(\lambda)$ crosses $S(\tau_0)$, (see
figure~\ref{fig:map}).  In doing so the entire curve $\mit\Upsilon$ is
mapped into a curve $\bar{\mit\Upsilon}$ in $S(\tau_0)$; this curve
has a tangent vector $\ell^\alpha$ which at every point of
$\bar{\mit\Upsilon}$ is numerically equal to the components of the
null vector $k^{\alpha}$ in the rest space of ${\pmb u}$ at the
corresponding point of $\mit\Upsilon(\lambda)$.

Neglecting the time variations of the metric and imposing $h_{0i}=0$
the geodesic equation for the light ray leads, after some algebra, to
the differential equations of $\bar{\mit\Upsilon}$. For the corresponding 
tangent field $\pmb{\ell}$ these differential equations
read:  
\begin{equation}
\label{eq:barelle}
\frac{d \bar{\ell}^{k}}{d \sigma}+ 
\bar{\ell}^{k} \left(\frac{1}{2}\bar{\ell}^{i}\partial_{i}h_{00}\right)+ 
\left(\partial_{i} h_{k j}-\frac{1}{2}\partial_{k} h_{ij}\right)
\bar{\ell}^{i}\bar{\ell}^{j}-\frac{1}{2}\partial_{k} h_{00}=0,
\end{equation}
where $\pmb{\bar{\ell}}$ is the unitary tangent vector field obtained
reparametrizing the tangent field $\pmb{\ell}$, which is in general
not unitary, and $\delta^{k s}$ is the Kronecker delta.

To integrate equation (\ref{eq:barelle}) one needs to fix boundary
conditions; these will be given by the components $\bar\ell^i_0$ of
the vector field $\bar\ell$ at the observation point, namely at
$\tau=\tau_0$ and $x^i=x^i_0$.  When the light ray is intercepted by
the satellite then the angles between the direction of the incoming
photon as {\it seen} by an observer comoving with the satellite and
the three spatial directions of a triad adapted to that observer, will
be the {\it observables}. These are essential to fix the required
boundary conditions namely to express the components $\bar\ell^i_0$ in
terms of known quantities alone. Our task now is to solve this
problem.

Let ${\bf u'}$ be the unitary vector field tangent to the satellite's
world line and let \{${\pmb \lambda_{\hat{a}}}$\} (where $\hat{a}= 1,
2, 3$) be a space-like triad carried by the satellite. The expressions
of the angle cosines ${\pmb e'}_{\hat a}$ are given by a well known
formula (see for instance \opencite{1990recm.book.....D} or
\opencite{1991ercm.book.....B}) containing the four-velocity of the
satellite, the components of the triad which identifies the
satellites's rest-frame, the metric coefficients at the observation
time and of course the components $\bar\ell_0^i$ which are our
unknowns.  The most important terms entering that formula are the
components of the satellite triad.  As well known the latter
identifies the rest-space of the observer comoving with the satellite.
Such a triad is defined up to spatial rotations, namely
transformations which leave the satellite's four-velocity $\pmb u'$
unchanged. Each triad vector ${\pmb\lambda_{\hat\alpha}}$ can be
expressed in terms of coordinate components with respect to the
coordinate basis $\{\pmb\partial_{\alpha}\}_{\alpha=\tau,x,y,z}$ in
the CM frame. If we denote the four-velocity of the satellite as
\footnote{The letter $s$ is for ``satellite''}
\begin{equation}
{\pmb u'}= \zeta \left( {\pmb \partial}_\tau + 
X_s {\pmb\partial}_x + Y_s {\pmb\partial}_y + Z_s {\pmb\partial}_z \right)
\end{equation}
where $\zeta$ is a function which makes the vector field $\pmb u'$ unitary and
$X_s,Y_s,Z_s$ are its $x,y,z$-components, a triad solution is given by 
\begin{eqnarray*}
{\pmb\lambda_{\hat{1}}}&=& T_1 {\pmb\partial}_\tau + X_1 {\pmb\partial}_x + 
Y_1 {\pmb\partial}_y  \\
{\pmb\lambda_{\hat{2}}}&=& T_2 {\pmb \partial_\tau} + X_2 {\pmb\partial}_x + 
Y_2 {\pmb\partial}_y + Z_2 {\pmb\partial}_z  \\
{\pmb\lambda_{\hat{3}}}&=& T_3{\pmb \partial_\tau} + Z_3 {\pmb\partial}_z.  
\end{eqnarray*}
where the $T_a,X_a,Y_a,Z_a$ are the $\tau, x,y,z$-components of the
corresponding $\pmb\lambda_{\hat a}$ vector of the triad and are well
defined functions of the components of the four-vector $\pmb u'$ and
of the metric coefficients, which are known quantities.  As stated, we
can relate the observed quantities ${\pmb e'}_{\hat{a}}$ to the
unknowns $\bar{\ell}^{\alpha}_{(0)}$ by means of the above properties
of the satellite frame.
 
Although the general expressions of the boundary conditions that we
have briefly mentioned will be published elsewhere, here we like to
show them in the simple case of a satellite moving on a circular orbit
around the Sun with coordinate angular velocity $\omega$. Recalling
that the origin of the coordinate system is at the center of mass of
the Solar System we obtain:
\begin{eqnarray*}
\bar{\ell}_{(o)}^{x} &=& \frac{1}{\sqrt{g_{xx}}\sqrt{\Sigma}}\times \\
                    &{}& \left[\frac{{\bf e'}_{\hat{1}}(x_0-x_\odot)
                         \sqrt{g_{yy}}\sqrt{-\Pi}+(y_0-y_\odot)
                         \sqrt{g_{xx}}(\omega\sqrt{\Sigma}+{\bf e'}_{\hat{2}} 
                         \sqrt{-g_{00}})}{({\bf e'}_{\hat{2}}\omega
                         \sqrt{\Sigma}-\sqrt{-g_{00}})}\right] \\
\bar{\ell}_{(o)}^{y} &=& \frac{1}{\sqrt{g_{yy}}\sqrt{\Sigma}}\times \\
                    &{}& \left[\frac{{\bf e'}_{\hat{1}}(y_0-y_\odot)
                         \sqrt{g_{xx}}\sqrt{-\Pi}-(x_0-x_\odot)
                         \sqrt{g_{yy}}(\omega\sqrt{\Sigma}+{\bf e'}_{\hat{2}} 
                         \sqrt{-g_{00}})}{({\bf e'}_{\hat{2}}\omega
                         \sqrt{\Sigma}-\sqrt{-g_{00}})}\right] \\
\bar{\ell}^{z}_{(o)} &=& -\frac{{\bf e'}_{\hat{3}}\,\sqrt{-\Pi}} 
                         {\sqrt{g_{zz}}\left({\bf e'}_{\hat{2}}\omega
                         \sqrt{\Sigma}-\sqrt{-g_{00}}\right)} 
\end{eqnarray*}
where $x_0, y_0, z_0$ are the coordinates of the satellite at the
observation time, $x_\odot, y_\odot, z_\odot$ are those of the Sun, 
$\Sigma\equiv g_{xx}(y_0-y_\odot)^2+g_{yy}(x_0-x_\odot)^2$, and $\Pi
\equiv g_{00}+(g_{xx}(y_0-y_\odot)^2+g_{yy}(x_0-x_\odot)^2)\omega^2$.

\section{Testing the model}
\label{sec:test}
The extension of the many-body model to higher orders is expected to
be complicated not only by the mathematical structure of the
relativistic equations, but also by the numerical methods needed to
implement it into a software code suitable for the reduction of the
satellite data when they become available.  Testing the model is of
course essential; a basic step is to verify whether to the order $
(v/c)^2 $ and under the same conditions our model is consistent with
the Schwarzschild non-perturbative model developed in
\citeauthor{1998AAp...332.1133D}
\shortcite{1998AAp...332.1133D,2001AAp...373..336D}. The tests that we
have devised had the purpose to verify whether: i) the perturbative
model restore full spherical symmetry when we consider the spherical
Sun as the only source of gravity; ii) the light deflection caused by
each individual body of the Solar System as evaluated in our
perturbative model agrees within the $(v/c)^2$ approximation with that
expected in the Schwarzschild metric under the same observational
conditions; iii) our model is able to reconstruct stellar distances.
The tests as to the point i) were completely successful hence we shall
only discuss those regarding points ii) and iii).

An analytical formula for the light deflection as a function of the
angular displacement $\psi$ of a star from the Sun, is given in
\inlinecite{1973grav.book.....M}.  Since that formula is itself
approximated to the order of $(v/c)^2$ and refers to a spherical Sun,
we expect that its predictions coincide with ours, in case the Sun is
the only source of gravity and spherical as well, at the $(v/c)^2$
order, that is at $\sim 0.1$~milliarcsec.  Taking the same stars and
computing the light deflection, the tests have shown a difference
between the values deduced from our model and those obtained with the
analytical formula of $ \lesssim 15~\mu $arcsec for almost limb
grazing light rays. The difference becomes rapidly less than $ 0.1~\mu
$arcsec for $ \psi>5^\circ $.  We are now performing a new series of
numerical tests with our perturbative model to better investigate the
reason of that discrepancy and the results will be published
separately as soon as they are available \cite{2003PhRvD..InPrep...D}.
 
In the exact Schwarzschild model we found a formula that expressed the
components of the tangent vector $ k^i $ to the null geodesic with
respect of the spatial polar axes $ \hat r $, $ \hat\theta $ and $
\hat\phi $ of a \textit{phase-locked} tetrad associated to an observer
moving on a circular orbit around the Sun.  Given the position of a
star and that of the observer (in Schwarzschild coordinates), we are
able to deduce the cartesian components of the tangent to the null
geodesic at the position of the observer by means of an almost
completely analytical procedure, because the complete set of
Schwarzschild coordinates of the star $(r^*,\theta^*,\phi^*) $ can be
recovered by analytical integration.  Practically, in this test we
have considered the observer at two opposite positions on his orbit
which we took as that of the Earth around the Sun (i.e. $
r_1=r_2=r_\oplus $, $ \theta_1=\theta_2=\pi/2 $ and $
\phi_2=\phi_1+\pi $). The stars are also lying on the orbital plane ($
\theta^*=\pi/2 $) so that the two observers are symmetrically placed
about the Sun ($ \phi^*=(\phi_1+\phi_2)/2 $).  The distance $r^*$ from
the stars ranges from $ 1 $~pc to $ 10 $~kpc.  With this configuration
it is easy to calculate the parallax of a star as $ p^*=1/r^* $~rad
(the distances being in AU), and so it is equally easy to determine
the numerical accuracy of the position in terms of angles by simply
subtracting the parallax $ p^*_s $ of the Schwarzschild model and the
approximate one reconstructed from the numerical integrations
described above, namely $ p^*_a $. The final tests show that the
difference $ p^*_s-p^*_a $ is $\sim 18~\mu $arcsec for the entire
distance range simulated.  Here again we find a difference perhaps due
to an insufficinet numerical accuracy and our present task is to
improve on this situation.

\nocite{2002PhRvD..65f4025K,1987thyg.book.....H}

\begin{acknowledgements}
  Work partially supported by the Italian Space Agency (ASI) under
  contracts ASI I/R/32/00 and ASI I/R/117/01, and by the Italian
  Ministry for Research (MIUR) through the COFIN 2001 program.  We
  thank the referee S.~A.~Klioner for his helpful comments and
  suggestions.
\end{acknowledgements}
\bibliography{mybibl}
\end{article}
\end{document}